\documentclass[12pt,english]{article}
\usepackage[longnamesfirst]{natbib}
\usepackage{url}
\makeatletter
\def\url@leostyle{%
    \def\UrlFont{\sf}}{\def\UrlFont{\small\ttfamily}}
\makeatother
\usepackage{geometry}
\usepackage{pstricks}

\usepackage[font=small,labelfont=bf]{caption}

\usepackage[hidelinks,breaklinks]{hyperref}
\usepackage[hyphenbreaks]{breakurl}

\usepackage{authblk}

\usepackage{amsmath}

\usepackage{amssymb}

\usepackage{nicefrac}

\newcommand{\ignore}[1]{}

\usepackage{color}

\usepackage[T1]{fontenc}
\usepackage{charter} % bitstream charter
\usepackage{microtype}

\usepackage{graphicx}

\usepackage[english]{babel}

\numberwithin{equation}{section}

\newcommand{\documentTitle}{Methodological Realism and Quantum Mechanics\thanks{This is a preprint of the following chapter: Michael E. Cuffaro, Methodological Realism and Quantum Mechanics, which has been submitted for publication in Lars-G\"oran Johansson and Jan Faye (eds.), \emph{How to Understand Quantum Mechanics -- 100 Years of Ongoing Interpretation}. It is the version of the author's manuscript prior to acceptance for publication. As such, this version \textbf{has not undergone editorial and/or peer review on behalf of the Publisher}. A link to the final authenticated version of the manuscript will be added here upon eventual publication.}}

\usepackage[yyyymmdd,hhmmss]{datetime}

\title{\documentTitle}
\author[*]{Michael E. Cuffaro}
\affil[*]{{\small Munich Center for Mathematical Philosophy,
    Ludwig-Maximilians-Universit\"at M\"unchen}}

\date{\today}

\begin{document}

\maketitle

%% no headers on the first page:
\thispagestyle{empty}

\begin{abstract}
  I distinguish two senses in which one can take a given physical theory to be `complete'. On the first, a complete physical theory is one that, in principle, completely describes physical reality. On the second, a complete physical theory is one that provides all of the conceptual resources one needs to describe any (in general probabilistic) physical phenomenon to any level of detail one likes, in principle. I argue that while the (neo-)Everettian approach to interpreting quantum mechanics aims to show that it is complete in the first sense, the (neo-)Bohrian approach begins from an understanding of quantum mechanics as being complete in the second sense. I then discuss some of the essential differences between how classical and quantum theory describe phenomena, and the way in which the quantum description can be thought of as a ``natural generalisation'' (to use Bohr's phrase) of the classical description. Finally, I elaborate upon the two visions of physics from which one can motivate the first and the second sense of completeness, respectively: \emph{metaphysical realism}, on the one hand, and what I will call \emph{methodological realism}, on the other---and discuss what one can say about the significance of the differences between quantum and classical description from each of these points of view. I suggest that there is a sense in which the views of (neo-)Everett and the views of (neo-)Bohr can be understood to be mutually supporting positions, from their respective perspectives, even though they are diametrically opposed.
\end{abstract}

% THANKS to: Laura. Simon Saunders, Michel Janssen. Jeff. Western. Berlin. Maybe thank Tom. Sebastien. Noah. Guido.

\section{Introduction}
\label{sec:intro}

In the opening paragraph of his paper, \emph{How to Teach Special Relativity}, John S. Bell wrote:

\begin{quote}
I have for long thought that if I had the opportunity to teach this subject, I would emphasize the continuity with earlier ideas. Usually it is the discontinuity which is stressed $\dots$ Often the result is to destroy completely the confidence of the student in perfectly sound and useful concepts already acquired. \citep[p.\ 67]{bell1976}.
\end{quote}

Although the topic of Bell's paper was special relativity, the attitude being expressed here is of course a very general one. In this chapter I will argue that one can approach quantum mechanics in a similar spirit, though perhaps not precisely in quite the same spirit as Bell might have wanted. But it is the spirit of what I will call the (neo-)Bohrian approach to the interpretation of quantum mechanics. Of which, I begin, in Section \ref{sec:completeness}, by distinguishing two senses in which one can take a given physical theory to be `complete'. On the first, a complete physical theory is a theory that, in principle, completely describes physical reality. On the second, one takes a complete physical theory to be one that provides all of the conceptual resources one needs to describe any (in general probabilistic) physical phenomenon to any level of detail one likes, in principle. I will argue that while the (neo-)Everettian approach to interpreting quantum mechanics aims to show that it is complete in the first sense, the (neo-)Bohrian approach begins from an understanding of quantum mechanics as being complete in the second sense, a sense of completeness that, like the first, can be motivated by centuries of successful physical research, and that, unlike the first, is secured by Gleason's theorem. In Section \ref{sec:conditions}, I will then elaborate upon the essential differences between the ways in which classical and quantum theory describe phenomena, and upon the sense in which the quantum description can be thought of as a ``natural generalisation'' (to use Bohr's phrase) of the classical description. In Section \ref{sec:realism} I will then elaborate upon the two visions of physics from which one can motivate the two senses of completeness that I discussed in Section \ref{sec:completeness}---what I will call \emph{metaphysical realism}, on the one hand, and what I will call \emph{methodological realism}, on the other---and discuss what one can say about the significance of the difference between quantum and classical description from each of these points of view. I will close the section by suggesting that there is a sense in which the views of (neo-)Everett and the views of (neo-)Bohr can be understood to be mutually supporting positions, from their respective perspectives, though they are diametrically opposed. Section \ref{sec:conclusion} summarises this chapter and concludes.

\section{Two senses of completeness}
\label{sec:completeness}

In the philosophical debates over the interpretation of quantum mechanics, one of the recurring points of contention has been whether the theory should be considered to be complete. This has been the case since at least as early as the well-known exchanges between Niels Bohr and Albert Einstein, with Einstein, famously, arguing against Bohr through a series of thought experiments relating to what we now call entangled systems (\citealt{bacciagaluppiCrullEPR}, ch. 1; \citealt{howard1985}; \citealt{demopoulosOnTheories}, ch.\ 4) that the theory is, in fact, incomplete and that it requires the specification of further parameters in order to unambiguously specify a system's physical state. Although, early on, most saw Bohr as the victor in these debates, the attitudes of a great many physicists and philosophers of physics have shifted remarkably in the intervening years. To be sure, Einstein's own statistical interpretation of the quantum state has its problems, as argued by \citet[]{bell1964} and others who have followed in the research programme that Bell started. Nevertheless, Einstein's contention that there is something missing in the quantum-mechanical description of a system is now firmly in the mainstream of the philosophy of quantum mechanics, although it is certainly not universal.

One of the dissenting voices against the shift in attitude has been the (neo-)Everett interpretation, which at the beginning of the so-called ``second quantum revolution'' \citep[]{aspectQuantumRevolutions} could be considered to be no more than a marginal position, but which has since, through the tireless efforts of its various defenders \citep[]{dewitt1973, saunders2010, vaidman2024}, come to also loom large in the mainstream of the interpretational debate. When confronted with a tension between the intuitive ideas with which we are familiar from our study of classical physics, on the one hand, and quantum mechanics, on the other, the (neo-)Everettian approach, to its great credit, chooses quantum mechanics, and maintains, against those who advocate for a hidden-variables approach, that quantum mechanics should be thought of as a complete theory, albeit with the caveat that it commits us to a radically different ontology than the one with which we were familiar prior to its discovery. As for Bohr's own view, it has been widely panned in the philosophical literature, though perhaps less so within the foundational physics literature. Long gone, in any case, are the days when Bohr's view, or anyway something or other that was claimed to be Bohr's view, represented the official or generally accepted way to interpret quantum mechanics.

By its many detractors, it has been decried as, at best, obscure, and at worst, inconsistent---his early defenders apparently, according to critics, having sided with Bohr largely out of personal loyalty to him rather than as a result of any deep engagement with his views. But whatever their actual motivations were, it is certainly true that the actual differences between Bohr's view and the views of many of those who took themselves to pursue physics in his spirit like Werner Heisenberg \citep[]{camilleri2009} did, over time, tend to get glossed over, in some cases causing further confusion regarding Bohr's distinctive position as a result (\citealt{howard2004}, \citealt{rynoCorrespondencePrinciple}, \citealt{toader2025}, ch.\ 3). E. T. Jaynes famously expressed his frustration with the general approach by describing it as ``a peculiar mixture describing in part realities of Nature, in part incomplete human information about Nature -- all scrambled up by Heisenberg and Bohr into an omelette that nobody has seen how to unscramble.'' \citep[]{jaynesOmelette}.

To be fair, historians of the philosophy of physics \citep[e.g.,][]{bacciagaluppi2015, camilleri2017, faye1991, faye2017a, folse1985, folse2017, howard1994b, howard2004, perovic2021}, despite some differences, have generally tended to be more charitable towards Bohr. Further, there have recently emerged a number of publications defending a so-called neo-Bohrian approach (see, especially, Bub, this volume;\nocite{bubNoQuantumWorld} \citealt{bub2017}, \citealt{cuffaroFeatureNotBug}; \citealt{demopoulosOnTheories}; \citealt{3m2020}; \citealt{JanasJanssenContextuality}; and the related view of \citealt{landsman2017}). The prefix, ``neo-'', sometimes indicated only in parentheses, is used to convey that although the underlying ideas are taken, essentially, to be Bohr's, much of the language used to convey them and many of the technical results that are appealed to in their defence would have been unfamiliar to Bohr (cf., Bub,\nocite{bubNoQuantumWorld} this volume, sec.\ 1). These include contemporary concepts of information, including so-called quantum information, as well as results, such as, seminally, Bell's derivations of what have since become known as the `Bell inequalities', and the wealth of further results that have ensued from the information-theoretic approach to foundations that Bell helped inspire. Part of my goal in this chapter will be to summarise my own take on these ideas, especially in connection with work with which I have been more closely linked.

(Neo-)Bohrian approaches aside, what is usually unrecognised, or at any rate unacknowledged in the debates between defenders of so-called hidden-variables approaches amongst each other, and in their debates with defenders of (neo-)Everettian approaches, is that there are in fact two different senses in which one can take a given physical theory to be complete. Most familiar to contemporary philosophers of physics, perhaps, is the sense of completeness implicit in Einstein's criticism that quantum mechanics is \emph{incomplete}. This being the idea that a complete physical theory is one that provides us, in principle, with a complete physical description of reality. From this point of view, arguably the only sense that can be made---if any can be made at all---of a view like Bohr's, and for that matter, similar views such as Heisenberg's, is that in claiming quantum mechanics to be complete, what they actually mean is that it is \emph{necessarily incompletable}---the justification, presumably, coming from results such as John von Neumann's infamous `no hidden-variables proof', which (as such) has since been discredited (\citealt{bell1966}; \citealt{hermann1933}; \citealt{hermann1935a}; \citealt{mermin1993}).\footnote{For further discussion, see \citet{acuna2021, acuna2021b, bub2010b, dieks2017, duncanJanssen2023}.}

There is another sense of completeness, however, that arguably better captures what Bohr and others meant when they claimed it for quantum mechanics. This is the idea that a complete theory is one that provides us with all of the conceptual resources we need to describe any given (in general, probabilistic) \emph{physical phenomenon} to any level of detail we like, in principle (cf., Bub, this volume,\nocite{bubNoQuantumWorld} sec.\ 2). As William Demopoulos has pointed out, that quantum mechanics satisfies this sense of completeness is a fact that has been rigorously proven by \citet{gleason1957}. It is not in doubt. Demopoulos writes:

\begin{quote}
[T]here is a concept of completeness that generalizes the classical concept and which was shown by Gleason to apply to the quantum theory. This concept does not require that an irreducibly statistical theory should derive its probability measures from a level of description that corresponds to Einstein’s real factual situations. Rather, completeness in the generalized sense established by Gleason requires that the quantum theory should generate all possible positive real-valued measures that are classical probability measures on Boolean subalgebras of the algebra of properties that the theory associates with a physical system. \citep[p.\ 179]{demopoulosOnTheories}.
\end{quote}

Although they predate Gleason's theorem, this is the sense of completeness that I and those who defend a similar view take to be implicit in Bohr's most important writings on the subject. Most, but not all, of these writings were published. One of the clearest expressions of the essence of what we take to be Bohr's view can be found in a letter, dated March 24, 1928, shortly after the completion of his so-called `Como paper' \citep[]{bohr1928},\footnote{The paper was originally presented at a conference in Como in the Italian province of Lombardy, honouring the centenary of Alessandro Volta's death, in 1927. For further discussion of the genesis of this paper, see \citet{degregorio2014}.} that he wrote to Paul Dirac.

\begin{quote}
  I quite appreciate your remarks that in dealing with observations we always witness through some permanent effects a choice of nature between the different possibilities. However, it appears to me that the permanency of results of measurements is \emph{inherent in the very idea} of observation; whether we have to do with marks on a photographic plate or with direct sensations the possibility of some kind of remembrance \emph{is of course the necessary condition for making any use} of observational results. It appears to me that the permanency of such results is \emph{the very essence of the ordinary causal space-time description}. This seems to me so clear that I have not made a special point of it in my article [= the Como paper]. What has been in my mind above all [, rather,] was the endeavour to represent the statistical quantum theoretical description \emph{as a natural generalisation} of the ordinary causal description and to analyze the reasons why such phrases like a choice of nature present themselves in the description of the actual situation. \citep[pp.\ 45--46, emphasis added]{bohrToDirac1928}.
\end{quote}

Notice how, for Bohr, (classical) concepts of observation and measurement constitute the starting point of his analysis of quantum theory, rather than an end toward which he is working. For Bohr, the ``ordinary causal description'' that such concepts enable is the necessary touchstone by which we are able to understand what quantum mechanics---insofar as it generalises that description in a particular way---is telling us about the world at all. Demopoulos elaborates:

\begin{quote}
  By the ``primacy of classical concepts'' for our understanding of quantum mechanics I mean---and I take Bohr to have meant---their primacy in the description of experimental results pertinent to the development and confirmation of the theory. \citep[p.\ 121]{demopoulosOnTheories}.
\end{quote}

Bohr's claim, in other words, is not that the ordinary causal description, refined by the concepts of classical physics, is to be thought of as primary with regard to the theoretical statements of whatever our final theory of physics turns out to be. Quantum theory itself, insofar as the dynamical phenomena it successfully describes are incapable of being fully captured in a satisfactory way by classical theory, or so one might argue, shows that such a claim must in fact be false. What Bohr means by the primacy of classical concepts is that they have \emph{evidentiary} primacy. Demopoulos goes on:

\begin{quote}
  Understood as a thesis about the epistemic framework within which physical theories are evaluated, the thesis of the primacy of classical concepts is entirely compatible with the idea that the principles and presuppositions of the classical framework are radically mistaken and incapable of providing an adequate theoretical basis for physics. \citep[p.\ 123]{demopoulosOnTheories}.
\end{quote}

Illustrating the idea through an example that he draws from the history of physics, Demopoulos considers the use made by Jean Perrin and J. J. Thomson of the relation known as \emph{Stoke's Law of Fall}, relating the drag force experienced by a particle, as it falls through a fluid medium, to its density and to the density and viscosity of the medium. As Demopoulos explains, Stoke's law, which is only valid for small spherical objects that can be assumed to produce negligible effects on the medium, guided Perrin's and Thomson's research into the properties of molecular and subatomic matter, despite the fact that it was thought unlikely that the relation expressed by the law held for the objects to which it was being applied.

\begin{quote}
  Nevertheless, Stoke's law isolates what, in Poincar\'e's terminology is a ``true'' relation---within a limited domain---between the rate of fall of spherical objects, density, and viscosity that is preserved under a change of application from the continuous media for which it was initially devised to discrete media. $\dots$ It illustrates the fact that the presuppositions of the principles which underlie an evidentiary framework might be false---and even \emph{known} to be false---and the principles themselves of only limited validity, without losing their effectiveness for probing the evidence for a theoretical claim, or refining the determination of a theoretical parameter. \citep[pp.\ 122--123]{demopoulosOnTheories}.
\end{quote}

The general idea, which Demopoulos ascribes to Bohr, that part of what it means to do science is to reconcile, and in that sense unify, various theoretical statements that are known to be valid within their respective, often very limited, domains, resonates with the way that another recent commentator on Bohr, Slobodan Perovi\'c, has, in a recent book,\footnote{For a review, see \citet[]{cuffaroPerovicReview}.} characterised the methodology that Bohr held to be central to his vision of physics. According to Perovi\'c, for Bohr, physical inquiry is to proceed in two stages, each associated with hypotheses of a distinct level of generality. The first stage of physical inquiry, for Bohr, involves the formation of what Perovi\'c calls concrete or ``low-level'' hypotheses and models whose validity is assumed not to reach beyond the particular experimental setups for which they were devised. This first stage is characterised by the use of everyday language, made precise using the mathematical concepts of classical physics, in which experimental particulars are observed and recorded \citep[p.\ 34]{perovic2021}. It results in what Perovi\'c calls an \emph{experimental account}, of how an object that is assumed, in accordance with our particular model, to be able to interact with our apparatus in a specific way, gives rise to the phenomena that we then observe \citep[p.\ 44]{perovic2021}.

In what Perovi\'c explains is Bohr's second stage of inquiry, by contrast, the object is to unify the various experimental accounts that have been produced in the first stage. Importantly, although everyday language does not directly constrain the ``master hypothesis'' that one aims to construct in this abstract stage, everyday language nevertheless indirectly constrains it insofar as the ultimate aim of this stage of inquiry is to obtain a comprehensive quantitative grasp of the overall experimental domain, in a way that preserves the validity that the lower-level models of experiments, constructed in the first stage, enjoy within their limited domains of applicability (\citealt{perovic2021}, pp.\ 50--51; cf. \citealt{rivatTheoreticalTerms}, sec.\ 3). Importantly, in order to ensure that an accepted master hypothesis continues to be open to revision in the light of new experiments, despite being implicit in one's accounts of the experimental data that is obtained thereby, one must take care to formulate a master hypothesis so that the novel theoretical relations that it reveals do not directly manifest themselves via controllable parameters that figure in our lower-level experimental accounts (\citealt{perovic2021}, pp.\ 139, 144). In quantum mechanics, of course, a fundamental way in which this is fulfilled is via the no-signalling principle (\citealt{bub2016}, pp.\ 51--52; \citealt{cuffaroInfCausality}, sec.\ 5--6; \citealt{demopoulosOnTheories}, pp.\ 164--165), which provides the guarantee that the novel nonlocal correlations described by the theory are transparent from a local point of view.

\section{The conditions of possible experience}
\label{sec:conditions}

In the third volume of their series on \emph{The Historical Development of Quantum Theory}, Jagdish Mehra and Helmut Rechenberg record the following excerpt from their conversations with Heisenberg:

\begin{quote}
  Heisenberg: ``[T]he fact that $XY$ was not equal to $YX$ was very disagreeable to me. I felt this was the only point of difficulty in the whole scheme, otherwise I would be perfectly happy.'' What this taught him, he continued: ``If one finds a difficulty in a calculation which is otherwise quite convincing, one should not push the difficulty away; one should rather try to make it the centre of the whole thing.'' \citep[p.\ 94]{mehraRechenbergVol3}.
\end{quote}

In classical mechanics, the commutativity of the algebra of observables associated with a physical system implies that once a state has been assigned to the system, any questions regarding the possible values that can be taken on by any of its observables is, in principle, already answered \citep[p.\ 61]{hughes1989}. Quantum mechanics is not like this. As Demopoulos explains, the Boolean algebras that can be used to represent the yes-or-no questions associated with a given observable do not combine into one globally Boolean algebra that can be used to represent them all.

\begin{quote}
  On the explication of \emph{classicality} that I believe is relevant to our understanding of quantum mechanics, the central characteristic of a framework or theory whose concepts are classical is the commutativity of the algebra of physical concepts---the parameters, physical magnitudes, and dynamical variables---with which it characterizes physical systems. Equivalently, classicality consists in the Boolean character of the algebra of all the properties or propositions that are associated with each physical system. On this view, classicality is a characteristic that attaches to the \emph{interrelations} of the physical concepts of a theory, rather than to the concepts themselves. \citep[pp.\ 126--127]{demopoulosOnTheories}.
\end{quote}

In his 1862 book \emph{On the Theory of Probabilities}, George Boole elucidated a number of what he called ``conditions of possible experience'' with respect to the observation of statistical data, such that ``When satisfied they indicate that the data \emph{may} have, when not satisfied they indicate that the data \emph{cannot} have resulted from an actual observation.'' (\citealt{boole1862}, p.\ 229, cited in \citealt{pitowsky1994}, p.\ 100). Boole explicates these conditions, which in general will vary from one experimental context to another, in the following way. Given a set of rational numbers, $p_1 \dots p_n$, representing the relative frequencies of $n$ logically connected events $E_1, \dots, E_n$, Boole asks: What are the necessary and sufficient conditions under which the $p_i$ can be realised as probabilities corresponding to the $E_i$ in some probability space? He then provides a general algorithm with which to answer the question. As \citet{pitowsky1994} has explained, essentially the idea is to begin by constructing a propositional truth table, such that the columns of the table correspond to the $E_i$, and such that each row is represented by a logically consistent vector of extremal (i.e., 0 or 1) probabilities corresponding to each.

\begin{figure}
  \begin{center}
    \mbox{} \\[12pt]
    \begin{tabular}{l l}
      \begin{tabular}{l | l | l | l | l}
        $E_1$ & $E_2$ & $E_3$ \\ \hline
        0   & 0  & 0 \\ \hline
        0   & 1  & 0 \\ \hline
        1   & 0  & 0 \\ \hline
        1   & 1  & 1
      \end{tabular}
      &
      \qquad\qquad
      \raisebox{-30pt}{
        \begin{pspicture}(1,1,1)
          \psset{linecolor=black,unit=3}
          \psline[](0,0,0)(1,1,1)
          \psline[](0,0,0)(0.35,0.65,0)
          \psline[](0.35,0.65,0)(1,1,1)
          \psline[](1,1,1)(0.5,0,0)
          \psline[](0,0,0)(0.5,0,0)
          \psline[linestyle=dotted](0.35,0.65,0)(0.5,0,0)
        \end{pspicture}
      }
    \end{tabular}
  \end{center}
  \caption{At the left: a propositional truth-table relating the extremal probabilities of three logically connected events. At the right, a schematic depiction of the polytope that is obtained upon taking convex combinations of the vectors represented by each row.}
  \label{fig:boole}
\end{figure}
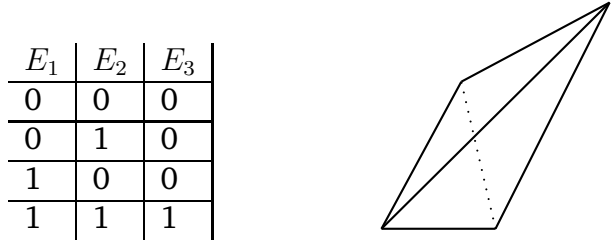

For instance, if our events are: $E_1$: it will snow in Montreal tomorrow, $E_2$: it will snow in Munich tomorrow, $E_3$: it will snow both in Montreal and in Munich tomorrow, then we will have the table depicted in Figure \ref{fig:boole}. If we now consider all possible convex combinations of the vectors, then the resulting probability space can be visualised as a 3-dimensional polytope, with each vertex representing one of the extremal vectors, and each facet associated with a linear inequality whose general form, for $n$ events, is given by:
\begin{align}
a_1p_1 + a_2p_2 + \dots + a_np_n + a \geq 0.
  \label{eqn:boole_inequality}
\end{align}
As \citet[]{pitowsky1994} showed, a special case of these conditions of possible experience, so-called by Boole, is given by the Bell inequalities.

\begin{figure}[h]
  \centering
  \includegraphics[scale=.85]{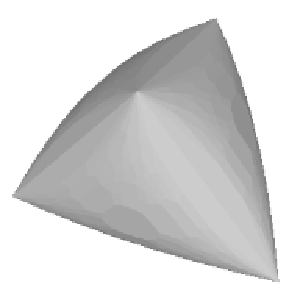} 
   \caption{A schematic visualisation of the elliptope describing the constraints on possible correlations between the three random variables described by Eq.\ \eqref{eq:genconst}.}
   \label{fig:elliptope}
\end{figure}

Drawing inspiration from Pitowsky's discussion and analysis of Boole, and from his further work on polytopes \citep[]{pitowsky1989}, \citet[ch.\ 3]{3m2020} considered more general constraints \emph{on correlations} between random variables, as described by a Pearson correlation coefficient: $\rho_{XY} := \frac{\langle XY \rangle}{\sigma_X\sigma_Y}$, where $X$ and $Y$ are random variables, $\langle XY \rangle$ is the covariance of $X$ and $Y$, and $\sigma_X$, $\sigma_Y$ are their standard deviations. In the context of a thought experiment proposed by N. David Mermin \citep[]{mermin1981}, in which two parties are given one of two correlated systems each and are each asked to measure their system using one of three possible settings, \citeauthor[]{3m2020} then showed that the correlations between the three balanced random variables,\footnote{It is assumed that possible values $x$ of $X$ are elements of a discrete set $\{ x_i \}$ or a continuous interval $[a, b]$ of real numbers (or of the union of such sets and intervals). A random variable $X$ is called \emph{balanced} if and only if: (1) whenever $x$ is a possible value, $\mathrm{-}x$ is a possible value; and (2) $x$ is as likely as $\mathrm{-x}$, i.e., $\mathrm{Pr}(x) = \mathrm{Pr}(\mathrm{-}x)$ (see \citealt{3m2020}, pp.\ 67--68).} $X$, $Y$, and $Z$, associated with each of these settings, is constrained by the following nonlinear inequality:
\begin{align}
  \label{eq:genconst}
  \quad 1 - \rho_{XY}^2 - \rho_{XZ}^2 - \rho_{YZ}^2 + 2 \, \rho_{XY} \, \rho_{XZ} \,
  \rho_{YZ} \ge 0.\quad
\end{align}
Geometrically, it describes an `inflated tetrahedron', or \emph{elliptope}, similar to the one schematically represented in Figure \ref{fig:elliptope}.

The elliptope and its underlying nonlinear inequality provide a useful visualisation of the logically possible correlations between the three random variables implicated in Mermin's thought experiment. Whether the full possibility space of the elliptope is actually saturated, however, depends, as \citeauthor{3m2020} discuss, on several additional factors. In a local hidden-variable model of the correlations, \citeauthor[]{3m2020} show that when the number of values per variable is limited to two (as it was in Mermin's original thought experiment), the possible correlations between these variables that can be realised are confined to a significantly limited subspace of the elliptope in the shape of a (non-inflated) tetrahedron. The reason for this is that, as they explain, modelling Eq.\ \eqref{eq:genconst} requires that one be able to define a joint probability distribution over the values of $X$, $Y$, and $Z$, which in a local hidden-variables theory---indeed in any theory that, like such a theory, has a commutative algebra of observables and is in that sense globally Boolean or classical---implies that the individual probability distributions over these values must also be well-defined. Better approximations of the elliptope can nevertheless be had, but they require more values per variable, such that saturation is achieved in the limit as the number of values per variable approaches infinity, though the correlation space thus delimited becomes increasingly difficult to compute.

When, instead, we model these correlations using quantum mechanics, it emerges that the key feature of the theory that allows us to saturate the entire correlation space of the elliptope is that, as \citet[]{3m2020} show, it allows one to assign a value to a sum of observable quantities---say, $\hat{S} \equiv \hat{S}_X + \hat{S}_Y + \hat{S}_Z$---without, in general, being required to assign values to the individual summands, $\hat{S}_X$, $\hat{S}_Y$ and $\hat{S}_Z$. \citet[]{3m2020} argue that this strongly suggests that the essential differences between classical and quantum theory are to be found in the analysis of their respective kinematics \citep[]{janssen2009}, i.e., in the constraints each imposes on the physical description of a system independently of the specifics of the system's dynamics.
% Regarding this, there are a number of further things that one can say. In classical mechanics, specifying the state, $\omega$, of a given physical system fixes the values of every observable quantity associated with it, from which it follows that the Boolean algebras, $\mathcal{A}, \mathcal{B}, \dots$ corresponding to the individual observables associated with the system are all embeddable into a single globally Boolean algebra. By contrast, in quantum mechanics, fixing the state, $| \psi \rangle$, only fixes the conditional probabilities, $\mathrm{Pr}(v_A|A), \mathrm{Pr}(v_B|B), \dots$ for obtaining a given value of an observable given an experiment on that observable.

As they then explain in more detail, although quantum mechanics is an irreducibly indeterministic theory, that in itself is not the most important difference between it and classical mechanics, for in the context of any given experiment on a system one can always treat the observed statistics, as one does in classical statistical mechanics, as arising from a prior probability distribution over the properties of the system that are relevant to that experimental context. Unlike classical theory, however, the Boolean algebras, $\mathcal{A}, \mathcal{B}, \dots$ of yes-or-no questions regarding the individual observables associated with a system, cannot generally be embedded into a single globally Boolean algebra, from which it follows that there is in general no global classical probability distribution over all of the system's observable parameters that one can use to predict the probability of a given experimental outcome irrespective of the experimental context. We can only say that conditional upon our inquiring about the observable $A$, there will be a particular probability distribution that one can use to characterise the possible answers to that question, which will in general be incompatible with the probability distribution that we should assign if we choose, instead, to measure the observable $B$. But if there is no globally Boolean algebra of observables, then representing the system's properties `as they are in themselves', i.e., irrespective of the way that one interacts with the system---even only probabilistically---becomes problematic. This difference between classical and quantum mechanics, with respect to the way that observable quantities are related by them, is striking. As for the Bohrian attitude toward it, it is summed up by Demopoulos, who, commenting on the difference, writes:

\begin{quote}
  The radical disparity between the algebraic structure of the classical and quantum-mechanical frameworks is not a problem that must be overcome, but is rather the true basis for the uniqueness of quantum mechanics in the evolution of physical theories that Bohr sought to highlight by his insistence on the methodological primacy of classical concepts. \citep[p.\ 135]{demopoulosOnTheories}
\end{quote}

\section{From metaphysical to methodological realism}
\label{sec:realism}

Recall that at the beginning of this chapter I distinguished between two senses of completeness. I mentioned that on the first, a complete physical theory is understood to be one in which it is possible in principle to completely describe physical reality, while on the second, a complete physical theory is understood to be one which has all of the conceptual resources needed to describe any, in general probabilistic, physical phenomenon to whatever level of detail we like, in principle. Neither of these senses of completeness is an arbitrary construct---both can be motivated by distinct conceptions of what it means to do physics.

I will begin with the first sense of completeness. Completeness in this sense is motivated by the worldview that I will call \emph{metaphysical realism}.\footnote{In \citet[]{cuffaroFeatureNotBug} I called this the ``traditional metaphysical picture.''} This is the idea, encapsulated in John Bell's famous dictum that ``Observables are \emph{made} out of beables'' \citep[p.\ 41, emphasis in original]{bell1973}, that the values of observable quantities should be understood to be the outward manifestations of a system's `ontic state', conceptualised as a set of physical properties that can be defined and reasoned about unambiguously, and in particular, without reference to the epistemic concepts of observation and measurement \citep[]{bell1989}. From this perspective, the idea that quantum mechanics can be a complete theory becomes problematic, for owing to the distinctive, non-Boolean, kinematical structure of the theory, the values of observable quantities cannot in general be straightforwardly interpreted, as they seemingly can in classical mechanics, as representing the antecedently given properties of a single physical system. This leaves anyone who continues to insist on a metaphysically realist conception of the goal of physics with essentially two options. We can posit further, perhaps unobservable, physical quantities from which the quantum-mechanical description can be derived or in some sense seen to follow. This is the route taken by the hidden-variables programme. The second option---the one taken by the (neo-)Everettian---is to argue that, appearances notwithstanding, quantum mechanics is, in fact, already complete in Bell's sense, with the qualification that the ontological structure of the world that it reveals is radically different than the one that we were seemingly being led to by classical physics.

As for the (neo-)Bohrian, the first thing to note is that unlike the picture of the vision of physics presupposed by the metaphysical realist, the picture of the vision of physics pursued by a (neo-)Bohrian---what I will call \emph{methodological realism}---is not one in which physical theories are expected to, in themselves, represent physical reality. For the (neo-)Bohrian, rather, physical theory, insofar as its goal is ultimately to represent known \emph{phenomena} in a systematic way, in that sense functions \emph{as a tool}---a tool, to be sure, that we make use of in order to represent the world (see \citealt{cuffaroFeatureNotBug}, sec.\ 3; cf. \citealt{leplinMethRealismSciRat}), but such a claim is distinct from the claim that a physical theory, in itself, \emph{is} such a representation (cf., \citealt{curiel2020}, secs.\ 2, 4; \citealt{doughertyRadicalEffectiveRealism}, sec.\ 6; \citealt{fayeClassicalIllusion}). What is exhibited by the quantum state, for a (neo-)Bohrian, is not, \emph{per se}, a collection of antecedently given properties possessed by a system. Rather, for a (neo-)Bohrian, the quantum state represents the structure of and interdependencies among the possible ways that one can effectively characterise a system in the context of a physical interaction. This is no less true of a classical state description \citep[sec.\ 3]{curiel2014}. But because the values of classical observables are in principle determinable independently of whether, or how, we interact with the system in order to assess them \citep[p.\ 61]{hughes1989}, we are invited to conceive of these values as somehow existing `in themselves', i.e., without reference to the concepts of observation and measurement in accordance with which we invariably determine them when we move from the armchair to the lab. By contrast, the more complex structure of observables described by quantum mechanics does not similarly invite us to conceive of the values of observable quantities in this way.

More usefully, in any given measurement context, quantum mechanics provides us with a recipe through which we can acquire information concerning a system through interactions with objects whose relevant parameters can, by assumption, be modelled as classical---an assumption justified, \emph{not by quantum mechanics}, but by hundreds of years of successfully employing classical physics at that scale.\footnote{Note that in general when I use the word `experiment' I mean this in a very broad sense. Any physical interaction through which I acquire information about a system, whether this happens inside or outside of the lab, can in principle count as an experiment on the system. If, for whatever reason, one believes that a given experiment is not capable of implementing the measurement of a particular parameter of interest, one can use quantum mechanics to express why and to what extent the phenomena it gives rise to depart from what we would ideally expect. Of course this presupposes that we are able to measure, precisely enough for the purpose, whatever parameters are able to provide us with information about the effectiveness of the experimental device when we `move back the cut' in this sense. The concept of `measurement' is ultimately epistemic. For further discussion, see Bub, this volume, sec.\ 2; \citet[sec.\ 6.5]{3m2020}.} The classical framework, in other words, is something that we already know to be able to provide us with a secure evidentiary basis from which to learn about the world, and through it we have been able to learn that the wider world is, in fact, most usefully described using the language of quantum mechanics.

In quantum mechanics, one does not presuppose that nature is such as to allow for a globally Boolean description of all of the dynamical physical phenomena that we are concerned to describe. It is not clear what the justification for such a belief could be. But that is not a problem for the methodological realist to solve, for as a vision of physics, methodological realism does not imply a commitment to metaphysical realism. It is strictly weaker. Moreover, it is in our view a more realistic conception of the way that science is actually done, and of how we have been able to uncover, through quantum mechanics and other modern physical theories, true empirical relations that obtain in the world \citep[cf.,][]{adlamModeratePhysicalPerspectivalism, weinbergerWilliamsWoodwardWorldlyInfrastructure}.

An analogy from the history of physics may help to drive home the latter point. In his book, \emph{Understanding Space-Time}, Robert DiSalle \citeyearpar[pp.\ 128--129]{disalle2006} explains how, in his investigations into the foundations of geometry, the neo-Kantian philosopher-scientist Hermann von Helmholtz was able to advance beyond the views of Immanuel Kant, which at the time had been experiencing a resurgence in interest \citep[see also][]{cuffaro19thCenturyScience}. Kant had argued that although non-Euclidean spatial relations are logically possible, it is in principle impossible to come to know them.

\begin{quote}
Thus in the concept of a figure that is enclosed between two straight lines there is no contradiction, for the concepts of two straight lines and their intersection contain no negation of a figure; rather the impossibility rests not on the concept in itself, but on its construction in space, i.e., on the conditions of space and its determinations; but these in turn have their objective reality, i.e., they pertain to possible things, because they contain in themselves \emph{a priori} the form of experience in general. \citep[A220--21 / B 268]{kant1781guyer}.
\end{quote}

In his penetrating critique of Kant, Helmholtz argued that Kant's analysis of our capacity to represent objects in space had been insufficiently fleshed out, and he maintained that it could be subjected to further analysis in the light of empirical research being done at the time, both by himself and by others, on the physiology of sense perception. Space, for Helmholtz, is a field that we construct to perceive things as we locally interact with our environment, the result of complex physiological processes that are describable in a lawlike way (\citealt[pp.\ 204]{beiser2014}; \citealt[pp.\ 85]{friedman2013}). Taking this perspective as his starting point, Helmholtz was able to show that the first four Euclidean postulates presuppose a more general constraint on the local constructibility of figures, the so-called principle of the free mobility of rigid bodies,\footnote{The first four postulates are: (I) that a straight line may be drawn between any two points; (II) that any given straight line segment may be extended indefinitely; (III) given a line segment, one may describe a circle whose centre is one of the segment's endpoints and whose radius is the length of the line; and (IV) that any two right angles are equal to one another.} or the assumption that moving a rigid body around in space does not deform it. By contrast, Euclid's fifth, the famous parallel postulate, which does not follow from the principle of free mobility, describes a global feature of space, namely that it conforms to a geometry in which the curvature is everywhere zero. Helmholtz then demonstrated how iterating a number of basic local constructions in accordance with postulates I--IV would result in a series of sense impressions through which one could in principle become acquainted with the global structure of a non-Euclidean space, along the way highlighting the importance of precisely defining the conditions of the possibility of measurement, in this case made more precise through the concept of a rigid body \citep[pp.\ 180, 201]{lenoir2006}.

From a (neo-)Bohrian point of view, the \emph{abstract} space of possibilities described by quantum mechanics is similarly pieced together from, in this case, the Boolean sub-algebras characterising the individual observables associated with a system, into a globally non-Boolean algebra \citep[cf.][pp.\ 129--131]{demopoulosOnTheories}. There is a temptation to, in some sense, reify this abstract structure \citep[cf.,][]{disalleSpacetimePhysicalGeometry}, an invitation which the (neo-)Everettian happily accepts, and on this basis claims that we have uncovered, through an analysis of the structure and presuppositions of quantum theory, a representation of the ontological structure of reality as it exists in itself, and that that structure is radically different than what we would have imagined had we not had need of quantum mechanics. Such a claim cannot be motivated for a (neo-)Bohrian, however, for its motivation does not come from empirical inquiry as such but from an \emph{a priori} conception of the goal of physics that, as a methodological but not a metaphysical realist, the (neo-)Bohrian is simply not committed to, and which must appear to her as an overreach.

The (neo-)Everettian and (neo-)Bohrian approaches to the interpretation of quantum mechanics are, in this sense, diametrically opposed. While, in accordance with the demands of metaphysical realism, the (neo-)Everettian attempts to derive or in some way argue that our classical experience emerges from an underlying quantum description which is taken to be complete in that sense, for the (neo-)Bohrian our effectively classically describable experience is itself the starting point for understanding the meaning of the quantum formalism. Yet despite this deep disagreement over the aims and methods of physics, in my final comment of this section I want to suggest that the connection between the (neo-)Everettian and (neo-)Bohrian views is more subtle and interesting than is often appreciated.

I begin with a superficial point. Unlike the approaches in the tradition of the hidden-variables programme, both the (neo-)Everettian and the (neo-)Bohrian have chosen to take what quantum theory is telling us about the world---the world \emph{underlying} our experience in the case of (neo-)Everett, or the world \emph{of} experience in the case of (neo-)Bohr---completely seriously and to trace out the consequences of what this means both for physics and for philosophy. The next point is one that I take to be less superficial. It is that there is a sense in which the (neo-)Bohrian and (neo-)Everettian viewpoints are mutually supporting. From the (neo-)Everettian point of view, although the (neo-)Bohrian approach is ultimately unsatisfactory---methodological realism does not, after all, imply metaphysical realism, and the latter is what is demanded---the fact that one can make sense of quantum mechanics from the (neo-)Bohrian perspective at all surely constitutes at least part of the empirical basis for believing in the world according to (neo-)Everett. While, from the (neo-)Bohrian perspective, the fact that it is possible to provide what Everett referred to as a ``metatheory of the observer'' \citep[p.\ 454]{everett1957} at all is, without doubt, of theoretical interest, regardless of what Everett himself took the physical significance of his metatheory to be. Moreover the (neo-)Everettian approach develops an intuitive sense in which the experiences of multiple observers can, from a logical point of view, be consistently described as all existing within a single closed system, even if there remains the further question, about which the (neo-)Bohrian and (neo-)Everettian fundamentally disagree, of what to make of the ontological status of the closed system that one describes in this way \citep[see][]{cuffaroHartmannOpenSystemsView}.

\section{Summary and conclusion}
\label{sec:conclusion}

I began, in Section \ref{sec:completeness}, by distinguishing two senses of completeness, which I associated with the (neo-)Everettian and (neo-)Bohrian approaches to quantum mechanics, respectively. In Section \ref{sec:conditions}, I then elaborated upon some of the essential differences between classical and quantum descriptions of phenomena, and upon one important sense in which quantum description naturally generalises classical description. Finally, in Section \ref{sec:realism} I elaborated upon the two visions of physics from which one can motivate the two senses of completeness I discussed in Section \ref{sec:completeness}, and we discussed what one can say about the significance of the difference between quantum and classical description from each  of these diametrically opposed points of view. I argued that there is a sense in which these two approaches are, despite their fundamental opposition, mutually supporting.

There is more to be said regarding the relation between (neo-)Bohr and (neo-)Everett, and about possible further points upon which they may agree \citep[][]{cuffaroHartmannOSVChapter}, but I will leave the further development of that question for another occasion. Instead I will close with the thought that there is a sense in which the (neo-)Everettian approach to the interpretation of quantum mechanics is a perfectly natural one, for anyone who is concerned to take what quantum mechanics is telling us about the word fully seriously. I believe that even a (neo-)Bohrian should be able to appreciate the power of Everett's idea. And yet we reject it. We follow Bohr and not Everett, not for physical reasons---for it is the very understanding of what it fundamentally means to do physics that is at stake---but for philosophical ones. Methodological realism begins from empiricism, from the idea that our theoretical knowledge of the world is ultimately something that we must learn through experience, and it maintains, in defiance of the metaphysical realist's demands, that there is no abstracting away from that, that would still count as theoretical knowledge by the standards we set. And yet, the very fact that we necessarily learn about the world through experience---or to be more precise, the abstract form of that fact \citep[]{cuffaroFeatureNotBug}---can nevertheless be the ground of a completely new kind of science, that allows us to easily express other kinds of facts, than the facts easily expressible under the assumption that the concepts of a theory always represent their objects in a globally Boolean way. This---the position that I have called methodological realism---represents, for us, the very idea of what it means \emph{to be} an empiricist. And we are at heart empiricists.

\section*{Acknowledgements}

Thanks to Jeff Bub for comments on a preliminary draft, and to Michel Janssen, Carlo Rovelli, and Simon Saunders for discussion.

\bibliographystyle{apa-good}
\bibliography{method_realism}{}

\end{document}